\documentstyle[aps,prb,twocolumn,floats,eqsecnum]{revtex}

\begin{document}

\newcommand{\gsim}{
\,\raisebox{0.35ex}{$>$}
\hspace{-1.7ex}\raisebox{-0.65ex}{$\sim$}\,
}

\newcommand{\lsim}{
\,\raisebox{0.35ex}{$<$}
\hspace{-1.7ex}\raisebox{-0.65ex}{$\sim$}\,
}

\bibliographystyle{prsty}

\title{
\begin{flushleft}
{\small 
PHYSICAL REVIEW B $\qquad$
\hfill
VOLUME {\normalsize 55}, NUMBER {\normalsize 5}
\hfill 
{\normalsize 1} FEBRUARY {\normalsize 1997-}I, {\normalsize 3050-3057}
}\\
\end{flushleft}  
Fokker-Planck and Landau-Lifshitz-Bloch equations for classical ferromagnets
}

\author{D.~A. Garanin
\renewcommand{\thefootnote}{\fnsymbol{footnote}}
\footnotemark[1]
}

\address{
I. Institut f\"ur Theoretische Physik, Universit\"at Hamburg,
Jungiusstr. 9, D-20355 Hamburg, Germany \\
\smallskip
{\rm(Received 9 July 1996)}
\bigskip\\
\parbox{14.2cm}
{\rm
A macroscopic equation of motion for the magnetization of a ferromagnet 
at elevated temperatures should contain both transverse and longitudinal 
relaxation terms and interpolate between Landau-Lifshitz equation at low 
temperatures and the Bloch equation at high temperatures.
It is shown that for the classical model where spin-bath interactions are 
described by stochastic Langevin fields and spin-spin interactions are 
treated within the mean-field approximation (MFA), such a 
``Landau-Lifshitz-Bloch'' (LLB) equation can be derived exactly from the 
Fokker-Planck equation, if the external conditions change slowly enough.
For weakly anisotropic ferromagnets within the MFA the LLB equation can be 
written in a macroscopic form based on the free-energy functional 
interpolating between the Landau free energy near $T_C$ and the 
``micromagnetic'' free energy,  which neglects changes of the magnetization 
magnitude $|{\bf M}|$, at low temperatures. 
[S0163-1829(97)03905-2]
\smallskip
\begin{flushleft}
PACS number(s): 75.40.Gb, 05.40.+j
\end{flushleft}
} 
} 

\maketitle

\renewcommand{\thefootnote}{\fnsymbol{footnote}}
\footnotetext[1]{ Electronic address: garanin@physnet.uni-hamburg.de }

\section{Introduction}

The famous Landau-Lifshitz equation, \cite{lanlif35} 
which is the basis of innumerable investigations of
magnetically ordered materials,
considers magnetization as a vector of fixed length and ignores its 
longitudinal relaxation.
Such an approach is obviously unsatisfactory at elevated temperatures since
magnetization is an average over some distribution function and its 
magnitude can change.
Alternatively, semiphenomenological ``soft-spin'' equations of motion 
for the spin density 
allowing for the longitudinal relaxation 
and for the influence of the bath described by stochastic Langevin terms 
are known in the theory of dynamic critical phenomena.
\cite{mamaz75,hohhal77,zin89}
A phenomenological deterministic equation of motion for 
the magnetization of magnetically ordered materials with the longitudinal 
relaxation terms, which is a direct generalization of the Landau-Lifshitz 
equation, was formulated by Bar'yakhtar \cite{bar84,bar87} and applied to 
the domain-wall dynamics at elevated temperatures. \cite{barivasaf89}
The Bar'yakhtar equation was conceived for the temperature range 
below the Curie point $T_C$; 
the theory does not answer what happens with phenomenological relaxation 
terms above $T_C$ and 
whether the Bloch equation used in the theory of EPR and NMR can be 
recovered in this region.

The simplest nontrivial model, 
for which the problem of finding an equation of 
motion for magnetization in the whole range of temperatures 
can be formulated,
is a semiphenomenological model considering
an isolated classical spin interacting with the bath modeled by 
stochastic Langevin fields.
The spin-spin interactions in this model, which lead to the 
ferromagnetism, can be taken into account on the next stage on the 
mean-field level.
Dynamics of such a spin is described by the Fokker-Planck equation (FPE), 
which can be solved analytically only in limiting cases, in particular, 
of low and high temperatures.
Reduction of the FPE using the modeling of the 
distribution function \cite{garishpan90} 
(the accuracy of this procedure was shown to be about 7\% in most 
situations) has led to the closed equation of 
motion for magnetization interpolating between the Landau-Lifshitz and 
Bloch equations at low and high temperatures --- the so-called 
``Landau-Lifshitz-Bloch'' (LLB) equation.
The LLB equation was also derived for a {\em quantum}
spin system interacting with a bath \cite{gar91llb} by the reduction 
of the density-matrix equation with the method similar to that used in 
the classical case.
A kind of LLB equation taking into account the spin-spin relaxation 
was obtained by Plefka \cite{ple90,ple93} for a 
quantum model with long-range ``spin-block'' interactions.

The coefficients in the relaxation terms of such a general LLB equation 
are nonlinear functions of magnetization itself; the only application of 
this equation up to now is that to the calculation of the nonlinear 
mobility of domain walls (DW) in rare-earth (RE) ferrites garnets, 
\cite{gar92zpb}
where the strongly thermally disordered spins of 
the RE sublattice do not interact with each other and are
subject to only 
the combined influence of the external field and the molecular field 
acting from the iron sublattice.
For the simplest one-sublattice weakly anisotropic ferromagnetic model 
below $T_C$ the dominant term in the molecular field is the homogeneous 
exchange, so that the directions of the molecular field and 
magnetization nearly coincide.
In this case the general LLB equation simplifies to its particular form 
similar to the Bar'yakhtar equation.
The latter was applied in Refs. \onlinecite{pangarrus90,gar91llb,gar91edw}
to calculate the domain-wall mobility in uniaxial 
ferromagnets in the whole temperature range and, in particular, near the 
phase transition from Bloch to linear (Ising-like) walls 
at some $T_B<T_C$ predicted by Bulaevskii and Ginzburg. \cite{bulgin64}
As this second-order phase transition is accompanied by changing 
the roles of transverse and longitudinal relaxation processes in the DW 
dynamics,
the DW mobility has a deep minimum at $T_B$.
\cite{pangarrus90,gar91llb,gar91edw}
This minimum, and thus the DW phase transition, was recently observed
in dynamic susceptibility experiments on Ba and Sr hexaferrites.
\cite{koeharjah93,koegarharjah93,harkoegar95}

An important dynamical scenario is that when the rate of changing of
magnetization (or of its spatial distribution), which can be controlled
by an external influence, is slow in comparison to the 
spin-relaxation rate.
This small parameter makes it possible to solve the Fokker-Planck 
equation exactly without making assumptions about the form of the 
distribution function.
For example, calculation of the low-frequency imaginary part of the 
longitudinal susceptibility leads to the exact analytical expression 
for the integral relaxation time $\tau_{\rm int}$, 
which is defined as the area under the magnetization relaxation curve 
after an abrupt infinitesimal change of the magnetic field. 
\cite{garishpan90,cofetal94,walkalcof94,cofcrokalwal95,gar96pre}
The quantity $\tau_{\rm int}$ describes, in particular, the thermoactivation 
escape rate of fine ferromagnetic particles with a uniaxial anisotropy 
over the potential barrier, which is valid, in contrast to the 
well-known Brown's solution, \cite{bro63} in the whole temperature range.
Such a situation is also characteristic for the dynamics of domain 
walls, whose velocity depends on the amplitude of the driving field and 
can be kept whatever small.
In this case the Fokker-Planck equation can be solved exactly, which 
leads to the exact form of the LLB equation, if the spin-spin 
interactions are considered within the mean-field approximation (MFA).
Derivation of this exact ``slow'' form of the LLB equation is the 
main purpose of this article.

The main part of the paper is organized as follows.
In Sec. \ref{secfpe} the Fokker-Planck equation for a classical spin,
its low- and high-temperature solutions, and the approximate 
reduction of the FPE to the Landau-Lifshitz-Bloch equation is outlined.
In Sec. \ref{secsol} the FPE is exactly solved in the slow-motion 
case and the slow LLB equation is derived.
In Sec. \ref{secfm} the simplified form of the latter 
for ferromagnets below and near $T_C$ is worked out.
In Sec. \ref{secdisc} further possible applications of the method and 
some unsolved problems are discussed.

\section{The Fokker-Planck and LLB equations}
\label{secfpe}

We shall describe a magnetic atom as a classical spin vector
${\bf s}$ of a unit length.
The magnetic and mechanical moments of the atom are given by 
$\bbox{\mu}=\mu_0 {\bf s}$ and ${\bf L}=\mu_0 {\bf s}/\gamma$, 
where $\gamma=ge/(2m_ec)$ is the gyromagnetic ratio. 
In the case of a weak coupling with the bath 
the dynamics of the vector ${\bf s}$ can be 
described with the help of the stochastic Landau-Lifshitz equation 
%
%
\begin{equation}\label{langeq}
{\bf \dot s} = \gamma [{\bf s}\times 
( {\bf H} + \bbox{\zeta} )] 
- \gamma\lambda [{\bf s}\times [{\bf s}\times {\bf H} ] ]
\end{equation}
with $\lambda\ll 1$, where correlators of the $\alpha,\beta=x,y,z$ 
components of the Langevin field $\bbox{\zeta}(t)$ are given by
%
%
\begin{equation}\label{zetacorr}
\langle \zeta_\alpha(t) \zeta_\beta(t') \rangle =
\frac{2\lambda T}{\gamma\mu_0} \delta_{\alpha\beta} \delta(t-t') .
\end{equation}

The Fokker-Planck equation corresponding to Eq. (\ref{langeq}) 
is formulated for the distribution function
$f({\bf N},t) = \langle \delta \bbox{(}{\bf N} - {\bf s}(t)\bbox{ )}\rangle$
on the sphere $|{\bf N}|=1$, where the average is taken over the 
realizations of $\bbox{\zeta}$.
Differentiating $f$ over $t$ with the use of Eq. (\ref{langeq}) and 
calculating the right part with the methods of stochastic theory
(see the Appendix), one comes to the Fokker-Planck equation 
\cite{garishpan90}
%
%
\begin{eqnarray}\label{fpe}
&&
\frac{\partial f}{\partial t} 
+ \frac{\partial}{\partial {\bf N}}
\bigg\{
\gamma \left [{\bf N}\times {\bf H} \right] 
- \gamma\lambda [{\bf N}\times [{\bf N}\times {\bf H} ] ]
\nonumber \\
&&
\qquad
{} + \frac{\gamma\lambda T}{\mu_0} 
\left [{\bf N}\times 
\left [{\bf N}\times \frac{\partial}{\partial {\bf N}} 
\right ] 
\right ]
\bigg\} f = 0 .
\end{eqnarray}
One can easily see that the distribution function
%
%
\begin{equation}\label{fequi}
f_0({\bf N}) \propto \exp[-{\cal H}({\bf N})/T],
\qquad {\cal H}({\bf s})=-\mu_0 {\bf Hs}
\end{equation}
satisfies (\ref{fpe}) at an equilibrium.
For the small coupling to the bath, $\lambda\ll 1$, Eq. (\ref{fpe}) 
coincides with the Fokker-Planck equation derived by Brown. \cite{bro63}

The equation of motion for the spin polarization (the first moment of 
the distribution function)
%
%
\begin{equation}\label{savr}
{\bf m}\equiv \langle {\bf s}\rangle = \int\!\! d^3N {\bf N}f({\bf N},t)
\end{equation}
of an assembly of magnetic atoms can be derived from Eq. (\ref{fpe}) 
and has the form
%
%
\begin{equation}\label{fmomeq}
{\bf \dot m} =
\gamma [{\bf m}\times {\bf H}] - \Lambda_N {\bf m}
- \gamma\lambda \langle [{\bf s}\times [{\bf s}\times {\bf H} ]]\rangle
\end{equation}
[cf. Eq. (\ref{langeq})], where $\Lambda_N$ is the characteristic 
diffusional relaxation rate or, for the thermoactivation escape problem,
the N\'{e}el attempt frequency given by
%
%
\begin{equation}\label{lamn}
\Lambda_N \equiv \tau_N^{-1} \equiv 2\gamma\lambda T/\mu_0 .
\end{equation}
It can be seen that Eq. (\ref{fmomeq}) is not closed but coupled 
to the second moments of the distribution function, 
$\langle s_i s_j \rangle$, in its last term. 
The behavior of Eq. (\ref{fmomeq}) is determined by the reduced
field $\xi_0$ given by
%
%
\begin{equation}\label{xi0def}
\xi_0 \equiv |\bbox{\xi}_0|, \qquad
\bbox{\xi}_0 \equiv \mu_0 {\bf H}/T .
\end{equation}
For $\xi_0\gg 1$ (low temperatures) the second term in Eq. (\ref{fmomeq}) can 
be neglected and the last term decouples for distribution functions 
localized about some direction: 
$\langle s_i s_j \rangle\cong m_i m_j$.
In this case the Landau-Lifshitz equation of the type (\ref{langeq}) 
for ${\bf m}$ without the stochastic field $\bbox{\zeta}$ is recovered.
In the high-temperature case, $\xi_0\ll 1$, the second term of Eq. 
(\ref{fmomeq}) dominates over the last one, which can be neglected.
Here one gets the equation of motion for ${\bf m}$ with the Bloch 
relaxation term.

In the intermediate region, $\xi_0\sim 1$, where 
the first-moment equation (\ref{fmomeq}) is not closed,
the resonance and relaxational behavior 
of the FPE (\ref{fpe}) is {\em not} described by Lorentz and Debye 
curves, and the deviations from the latter reach 7\% at 
$\xi_0\sim 3$. \cite{garishpan90}
Neglecting these features, one can obtain an {\em isolated} 
equation of motion 
for the spin polarization of an assembly of magnetic atoms  
choosing the distribution function in a form 
\cite{garishpan90} 
%
%
\begin{equation}\label{fmod}
f({\bf N},t) = \frac{\exp[\bbox{\xi}(t){\bf N}]}{{\cal Z}(\xi)},
\qquad {\cal Z} = 4\pi \frac{\sinh\xi}{\xi}
\end{equation}
[cf. Eqs. (\ref{fequi}) and (\ref{xi0def})],
where $\bbox{\xi}(t)$ is chosen so that the first moment equation 
(\ref{fmomeq}) is satisfied.
Calculating the terms of Eq. (\ref{fmomeq})
with the help of Eq. (\ref{fmod}), one arrives \cite{garishpan90}
at the LLB equation for the 
nonequilibrium reduced field $\bbox{\xi}(t)$ 
%
%
\begin{equation}\label{llbxi1}
%
\bbox{\dot\xi} = \gamma [ \bbox{\xi} \times {\bf H} ] 
- \Gamma_1 
\left( 1 - \frac{ \bbox{\xi\xi}_0 }{\xi^2} \right)\bbox{\xi}
- \Gamma_2
\frac{ [\bbox{\xi}\times [\bbox{\xi}\times \bbox{\xi}_0 ] ]  }{\xi^2} 
\end{equation}
with the longitudinal and transverse relaxation rates 
%
%
\begin{equation}\label{G1G2}
\Gamma_1 = \Lambda_N \frac{B(\xi)}{\xi B'(\xi)},
\qquad 
\Gamma_2 = \frac{\Lambda_N}{2} \left( \frac{\xi}{B(\xi)} - 1 \right) ,
\end{equation}
where $\Lambda_N$ is given by Eq. (\ref{lamn}), 
$B(\xi) = \coth\xi - 1/\xi$ is the Langevin function and 
$B'(\xi) \equiv dB(\xi)/d\xi$.
The asymptotic forms of $\Gamma_1$ and $\Gamma_2$ are given by
%
%
\begin{equation}\label{G1asy}
\Gamma_1 \cong
\left\{
\begin{array}{ll}
\displaystyle
\Lambda_N \left( 1 + \frac{2}{15} \xi^2 \right),
& \xi \ll 1 \\
\displaystyle
\Lambda_N \xi \left( 1 - \frac{1}{\xi} \right),
& \xi \gg 1 ,
\end{array}
\right.
\end{equation}
and
%
%
\begin{equation}\label{G2asy}
\Gamma_2 \cong
\left\{
\begin{array}{ll}
\displaystyle
\Lambda_N \left( 1 + \frac{1}{10} \xi^2 \right),
& \xi \ll 1 \\
\displaystyle
\frac{1}{2} \Lambda_N \xi \left( 1 + \frac{1}{\xi^2} \right),
& \xi \gg 1 ,
\end{array}
\right.
\end{equation}
The relaxation rates of such a type also appear as a result of
calculation of the high-frequency longitudinal susceptibility and
the far-from-resonance transverse one. \cite{garishpan90}
The quantity $\Gamma_1$ is also proportional to the ``effective
eigenvalue'' $\lambda_{\rm ef}$ of Ref. \onlinecite{cofetal94}.   
One can see that the equilibrium solution of Eq. (\ref{llbxi1}) is 
$\bbox{\xi}=\bbox{\xi}_0$.
The nonequilibrium spin polarization ${\bf m}$ is given by 
%
%
\begin{equation}\label{mxi}
{\bf m} = m \,\bbox{\xi}/\xi,  \qquad m = B(\xi) .
\end{equation}
The LLB equation for $\bbox{\xi}$, Eq. (\ref{llbxi1}), can be written 
in the alternative equivalent form
%
%
\begin{equation}\label{llbxi2}
\bbox{\dot\xi} = \gamma [\bbox{\xi}\times {\bf H} ] 
- \Gamma_1 ( \bbox{\xi} - \bbox{\xi}_0 )
- (\Gamma_2 - \Gamma_1)
\frac{ [\bbox{\xi}\times [\bbox{\xi}\times \bbox{\xi}_0 ] ]  }{\xi^2} .
\end{equation}
Here it can be seen that in the high-temperature region,
$\xi, \xi_0 \ll 1$, where $B(\xi)\cong \xi/3$ and
$\Gamma_1 \cong \Gamma_2 \cong \Lambda_N$,
the Landau-Lifshitz double-vector-product relaxation term becomes small
and the Bloch equation is recovered.
On the other hand, at low temperatures, when $\xi, \xi_0 \gg 1$,
the magnitude of the vector ${\bf m}$ in Eq. (\ref{mxi})
saturates in most situations
at $m = B(\xi)\cong 1$, and the longitudinal relaxation term in Eq. 
(\ref{llbxi1}) no longer plays a role.
Here the usual Landau-Lifshitz equation is recovered.
Using Eq. (\ref{mxi}) one can derive the LLB equation for the spin 
polarization ${\bf m}$ itself.
The result can be written as
%
%
\begin{eqnarray}\label{llbm1}
&&
{\bf \dot m} = 
\gamma [{\bf m}\times {\bf H}]
- \Lambda_N \left( 1 - \frac{{\bf m}\bbox{\xi}_0}{m\xi} \right){\bf m}
\nonumber
\\
&& \qquad
{} - \gamma\lambda \left( 1 - \frac{m}{\xi} \right) 
\frac{ [{\bf m}\times [{\bf m}\times {\bf H} ]] }{ m^2 } 
\end{eqnarray}
[cf. Eqs. (\ref{langeq}) and (\ref{fmomeq})], where 
$\xi=\xi(m)$ is determined implicitly by the relation $m=B(\xi)$.
Note that here at low temperatures, $\xi\gg 1$,
the coefficient before the transverse  
relaxation term goes to $\gamma\lambda$, whereas 
the longitudinal one is nonessential, if $m$ is saturated.
At high temperatures, $\xi\ll 1$, the relaxation term in Eq. (\ref{llbm1})
acquires the Bloch form $\Lambda_N ( {\bf m} - {\bf m}_0 )$ 
with ${\bf m}_0\cong \bbox{\xi}_0/3$ [see also Eq. (\ref{llbxi2})].
The quantum generalization of the classical LLB equation written above was
given in Ref. \onlinecite{gar91llb}.
The latter was applied in Ref. \onlinecite{gar92zpb} to study the nonlinear 
dynamics 
of the RE sublattice of rare-earth ferrites garnets near the 
magnetization compensation point.

For small deviations from equilibrium, where 
$\bbox{\xi}\cong\bbox{\xi}_0$ and, accordingly, 
${\bf m}\cong {\bf m}_0\equiv B(\xi_0)\bbox{\xi}_0/\xi_0$, 
one can put the LLB equation (\ref{llbm1}) 
[or, more conveniently, directly Eq. (\ref{llbxi1})]
into the form
%
%
\begin{eqnarray}\label{llbm2}
&&
{\bf \dot m} = 
\gamma [{\bf m}\times {\bf H}]
- \Gamma_1 
\left( 1 - \frac{ {\bf mm}_0 }{m^2} \right){\bf m}
\nonumber\\
&&\qquad
{} - \Gamma_2
\frac{ [{\bf m}\times [{\bf m}\times {\bf m}_0 ] ] }{m^2} , 
\end{eqnarray}
where the relaxation frequencies $\Gamma_1$ and $\Gamma_2$ are functions 
of $\xi_0$.
A kind of LLB equation similar to Eq. (\ref{llbm2}) was obtained 
by Gekht {\em et al.}, \cite{gekignraishl76} who assumed, 
for the calculation of the linear transverse 
dynamic susceptibility,
instead of Eq. (\ref{fmod}) a distribution function of the form
$f({\bf N},t) = f_0({\bf N}) [1+\bbox{\alpha}(t){\bf N}]$, 
where $f_0$ is given by Eq. (\ref{fequi}) and $\bbox{\alpha}$ corresponds to 
$\bbox{\xi}-\bbox{\xi}_0$ in our notations.
Although Gekht {\em et al.} claimed that ``the single-moment approximation 
is permissible for small deviations from equilibrium,'' Eq. (\ref{llbm2})
is in fact only approximate, as well as the more general Eq. (\ref{llbm1}).
The latter, in contrast, can be applied and has a rather good accuracy in 
situations where deviations from equilibrium are large, as was checked in
Ref. \onlinecite{garishpan90}.
In Sec. III we will consider the solution of the FPE (\ref{fpe})
for slowly varying field ${\bf H}(t)$.
In this case the deviations from the instantaneous equilibrium state are 
small and the FPE can be solved exactly without assumptions about the 
form of the distribution function $f({\bf N},t)$.

\section{The ``slow'' LLB equation}
\label{secsol}

If the magnetic field ${\bf H}$ slowly changes its magnitude and 
direction, the solution of the Fokker-Planck equation (\ref{fpe}) 
slightly deviates from the instantaneous equilibrium one and can be searched 
for in the form
%
%
\begin{equation}\label{fslow}
f({\bf N},t) \cong \frac{\exp[\bbox{\xi}_0(t){\bf N}]}{{\cal Z}(\xi_0)}
[1 + Q({\bf N},t)], 
\qquad Q \ll 1 ,
\end{equation}
where $\bbox{\xi}_0(t)\equiv \mu_0 {\bf H}(t)/T$.
The correction function $Q({\bf N},t)\propto |{\bf \dot H}|$ and, 
additionally, it depends slowly on time, so that 
$\dot Q \propto |{\bf \dot H}|^2$.
Neglecting this small term, one obtains from Eq. (\ref{fpe}) the equation 
for $Q$ having the form
%
%
\begin{eqnarray}\label{qeq}
&&
[{\bf N}\times \bbox{\xi}_0 ]\frac{\partial Q}{\partial {\bf N}} 
+ \lambda \left( \frac{\partial}{\partial {\bf N}} + \bbox{\xi}_0 \right)
\left [{\bf N}\times 
\left [{\bf N}\times \frac{\partial Q}{\partial {\bf N}} 
\right ] 
\right ]
\nonumber\\
&&\qquad
= \tau_0 ( {\bf m}_0 - {\bf N} ) \bbox{\dot\xi}_0 ,
\qquad {\bf m}_0\equiv B(\xi_0)\frac{ {\bf H} }{ H } ,
\end{eqnarray}
where $\tau_0\equiv \mu_0/(\gamma T)$.
One can see that in leading order the correction $Q({\bf N},t)$ 
is determined by the instantaneous values of the 
magnetic field ${\bf H}(t)$ and its first derivative ${\bf \dot H}$.
The right-hand part of this equation can be separated into the terms 
describing the temporal changes of the magnitude and of the direction of 
${\bf H}$ as
%
%
\begin{equation}\label{rhssep}
( {\bf m}_0 - {\bf N} ) \bbox{\dot\xi}_0 =
 {\bf N}[\bbox{\xi}_0 \times \bbox{\Omega} ] 
+ \left[ m_0 - \frac{ {\bf N}\bbox{\xi}_0 }{ \xi_0 } \right]
{\bf \dot\xi}_0
\end{equation}
where 
%
%
\begin{equation}\label{omega}
\bbox{\Omega}\equiv 
 [\bbox{\xi}_0 \times \bbox{\dot\xi}_0 ]/ \xi_0^2 
\end{equation}
is the precession frequency of the vector $\bbox{\xi}_0$.
In the spherical coordinate system with $z$ axis along $\bbox{\xi}_0$ 
Eq. (\ref{qeq}) for $Q(x,\varphi)$, where $x\equiv\cos\theta$, takes on the 
form 
%
%
\begin{eqnarray}\label{qeqsph}
&&
\xi_0\frac{\partial Q}{\partial\varphi}
+ \lambda 
\left\{
\left( \frac{\partial }{\partial x} + \xi_0 \right)
(1-x^2)\frac{\partial }{\partial x}
+ \frac{1}{1-x^2} \frac{\partial^2 }{\partial \varphi^2}
\right\}Q
\nonumber\\
&&\qquad
= \tau_0\xi_0\sqrt{1-x^2} [ \Omega_y\cos\varphi - \Omega_x\sin\varphi ]
\\
&&\qquad\qquad\qquad
{} + \tau_0 (x - m_0) {\bf \dot\xi}_0 ,
\nonumber
\end{eqnarray}
where $\Omega_x$ and $\Omega_y$ are
$x$ and $y$ components of the vector $\bbox{\Omega}$.

The solution of the linear differential equation (\ref{qeqsph}) is a sum 
of two contributions induced by the transverse and longitudinal 
inhomogeneous terms: $Q=Q_\perp + Q_\|$.
Using the substitution 
%
%
\begin{equation}\label{qsubst}
Q_\perp = Q_x \cos\varphi + Q_y \sin\varphi, 
\qquad Q_+ \equiv Q_x + i Q_y,
\end{equation}
one comes to the equation 
%
%
\begin{eqnarray}\label{q+eq}
&&
Q_+
+ \frac{ i \lambda }{ \xi_0 }
\left\{
\left( \frac{d }{dx} + \xi_0 \right)
(1-x^2)\frac{d }{dx}
- \frac{1}{1-x^2} 
\right\}Q_+
\nonumber\\
&&\qquad\qquad
= \tau_0\Omega_+ \sqrt{1-x^2} ,
\end{eqnarray}
where $\Omega_+ \equiv \Omega_1 + i \Omega_2$.
This equation cannot in general be solved analytically, 
but the latter is possible in the typical case 
of the weak coupling to the bath, $\lambda\ll 1$.
For $\lambda/\xi_0 \ll 1$ one can easily find the 
solution iteratively, which yields
%
%
\begin{equation}\label{q+sol1}
Q_+ \cong \tau_0\Omega_+ \sqrt{1-x^2}
\left[ 1 + \frac{ i \lambda }{ \xi_0 } (2 + \xi_0 x) + \ldots \right].
\end{equation}
On the other hand, in the high-temperature region, where $\xi_0 \ll 1$,
one can neglect $\xi_0$ in the round brackets in Eq. (\ref{q+eq}), 
after which Eq. (\ref{q+eq}) can be analyticaly solved to yield
%
%
\begin{equation}\label{q+sol2}
Q_+ \cong \tau_0\Omega_+ \sqrt{1-x^2}
\frac{ 1 + 2i \lambda/\xi_0 }{ 1 + (2\lambda/\xi_0)^2 } .
\end{equation}
These two solutions overlap in the region $\lambda \ll \xi_0 \ll 1$, 
and thus they can be sewn together in the whole range of temperatures
into the formula, which can be obtained by replacing the numerator of the 
fraction in (\ref{q+sol2}) by $1 + (i \lambda/\xi_0) (2 + \xi_0 x)$.

The equation for $Q_\|(x)$ can be written as
%
%
\begin{equation}\label{qpareq}
\left( \frac{d }{dx} + \xi_0 \right)
(1-x^2)\frac{dQ_\|}{dx} =
\Lambda_N^{-1}(x - m_0) {\bf \dot\xi}_0 .
\end{equation}
It can be solved in two steps with the help of the substitution
$P(x) \equiv (1-x^2) dQ_\|/dx$.
First, integrating Eq. (\ref{qpareq}) one gets
%
%
\begin{equation}\label{psol}
P(x) = \frac{{\bf \dot\xi}_0 }{ \Lambda_N\xi_0 }
\left[ x - \coth\xi_0 + \frac{ e^{-\xi_0 x} }{ \sinh\xi_0 } \right] .
\end{equation}
Then, $Q_\|$ is given by 
%
%
\begin{equation}\label{qparsol}
Q_\|(x) = \int_{-1}^x\! \frac{ dx' }{ 1-x'^2 } P(x') + C ,
\end{equation}
where the constant $C$ is determined from the normalization condition 
for the distribution function (\ref{fslow}).

Now, the function $Q({\bf N},t)$ having been 
determined, one can calculate the 
spin polarization ${\bf m}$ using Eqs. (\ref{savr}) and (\ref{fslow}).
Returning to vector designations, one comes to the result 
%
%
\begin{eqnarray}\label{msol}
&&
{\bf m} \cong B(\xi_0) 
\left\{
\left(
1 + \frac{ \xi_0 B' }{ \Gamma_{1,{\rm int}}B } 
\frac{ {\bf H\dot H} }{ H^2 }
\right) \frac{ {\bf H} }{ H }
+ \frac{ \gamma H }{ (\gamma H)^2 + \Gamma_2^2 } 
\right.
\nonumber\\
&&\qquad
\left.
{} \times
\left(
\frac{ [{\bf H}\times {\bf \dot H}] }{ H^2 }
+ \frac{ \Gamma_2 }{ \gamma H }
\frac{ [{\bf H}\times [{\bf H}\times {\bf \dot H}]] }{ H^3 } 
\right)
\right\} ,
\end{eqnarray}
where $\Gamma_2$ is the transverse relaxation rate given by Eq. (\ref{G1G2})
and $\Gamma_{1,{\rm int}}$ is the inverse of the integral longitudinal 
relaxation time 
$\tau_{\rm int}$,
%
%
\begin{eqnarray}\label{tauint}
&&
\frac{ 1 }{ \Gamma_{1,{\rm int}}} \equiv \tau_{\rm int} = 
\frac{ 1 }{ \Lambda_N \xi_0\sinh\xi_0 B'(\xi_0) }
\int_{-1}^1\! dx \frac{ e^{\xi_0 x} }{ 1-x^2 }
\nonumber\\
&&\qquad
{} \times  
\left[ x - \coth\xi_0 + \frac{ e^{-\xi_0 x} }{ \sinh\xi_0 } \right]^2 ,
\end{eqnarray}
which is determined as the area under the magnetization relaxation curve 
after an abrupt infinitesimal change of the longitudinal magnetic field.
\cite{garishpan90,gar96pre}
Equation (\ref{msol}) describes the lagging of the spin polarization 
${\bf m}$ from its quasiequilibrium value ${\bf m}_0(t)$ of Eq. (\ref{qeq}), 
which is determined by the small derivative ${\bf \dot H}(t)$.
The asymptotic forms of $\Gamma_{1,{\rm int}}$ in Eq. (\ref{tauint}) read
%
%
\begin{equation}\label{G1intasy}
\Gamma_{1,{\rm int}}\cong
\left\{
\begin{array}{ll}
\displaystyle
\Lambda_N \left( 1 + \frac{1}{9} \xi_0^2 \right),
& \xi_0 \ll 1 \\
\displaystyle
\Lambda_N \xi_0 \left( 1 - \frac{1}{\xi_0} \right),
& \xi_0 \gg 1 .
\end{array}
\right.
\end{equation}
Comparing Eqs. (\ref{G1intasy}) and (\ref{G1asy}) one can see that 
$\Gamma_1 > \Gamma_{1,{\rm int}}$.
The relative deviation 
$\delta \equiv \Gamma_1/\Gamma_{1,{\rm int}}- 1$ attains a value 
$\delta\approx 0.07$ 
at $\xi_0\approx 3$. \cite{garishpan90}

The next problem is to write down the equation of motion for ${\bf m}$, 
which has the solution (\ref{msol}).
It is especially important if the spin-spin interactions are taken into 
account within the MFA (see the next section).
In this case ${\bf H}$ is replaced by the molecular field $H^{\rm MFA}$ 
containing 
${\bf m}$ itself, and Eq. (\ref{msol}) is in fact a differential equation 
for ${\bf \dot m}$, which should be still simplified.
It can be done differentiating Eq. (\ref{msol}) over time and neglecting terms 
of order ${\bf \dot {\it H}}^2 $ coming from the correction terms 
with ${\bf \dot H}$ in Eq. (\ref{msol}).
This leads to
%
%
\begin{equation}\label{mdot}
{\bf \dot m} \cong 
\xi_0 B'(\xi_0)  
\frac{ ({\bf H\dot H}){\bf H} }{ H^3 }
- B(\xi_0)
\frac{ [{\bf H}\times [{\bf H}\times {\bf \dot H}]] }{ H^3 } . 
\end{equation}
Now ${\bf \dot H}$ in this relation should be expressed through 
${\bf m}$ with the help of Eq. (\ref{msol}), which after some vector algebra 
leads to the ``slow'' LLB equation 
%
%
\begin{eqnarray}\label{llbs}
&&
{\bf \dot m} = 
\gamma [{\bf m}\times {\bf H}]
- \Gamma_{1,{\rm int}}
\left( 1 - \frac{ {\bf mm}_0 }{m^2} \right){\bf m}
\nonumber\\
&&\qquad
{} - \Gamma_2
\frac{ [{\bf m}\times [{\bf m}\times {\bf m}_0 ] ] }{m^2} , 
\end{eqnarray}
where ${\bf m_0}$ is given by Eq. (\ref{qeq}) and which is the refinement of 
Eq. (\ref{llbm2}) in the slow-motion situation.
The quantities $\Gamma_1$ of Eq. (\ref{G1G2}) and 
$\Gamma_{1,{\rm int}}$ of Eq. (\ref{tauint}) have the same leading
high- and low-temperature asymptotes, and, as was said above, they differ 
by no more than 7\% in the whole range of temperatures.  
The same order of magnitude also characterizes the difference between 
the Debye one-relaxator form of the longitudinal dynamic susceptibility 
$\chi_\|(\omega)$ following from Eq. (\ref{llbm2}) and the actual form
of $\chi_\|(\omega)$ 
following from the solition of the exact Fokker-Planck equation (\ref{fpe}) 
at intermediate temperatures. \cite{garishpan90}
It should be noted that in the fast-motion situations equation 
(\ref{llbm2}) is better than Eq. (\ref{llbs}), since it yields the exact 
leading (imaginary) term of the high-frequency expansion of 
$\chi_\|(\omega)$. \cite{garishpan90}

\section{LLB equation for ferromagnets}
\label{secfm}

For definiteness we consider the classical ferromagnetic model with the 
biaxially anisotropic exchange interaction
%
%
\begin{eqnarray}\label{fham}
&&
{\cal H} = -\mu_0 \sum_i {\bf H}_i {\bf s}_i
- \frac{1}{2} \sum_{ij} J_{ij} 
( \eta_x s_{xi}s_{xj} 
\nonumber\\
&&\qquad\qquad
{} + \eta_y s_{yi}s_{yj} + s_{zi}s_{zj} ) , 
\end{eqnarray}
where $\eta_x \leq \eta_y \leq 1$ are the anisotropy coefficients.
The dynamics of this model interacting with the bath is described by the
stochastic Landau-Lifshitz equation 
%
%
\begin{equation}\label{fmlleq}
{\bf \dot s}_i = \gamma [{\bf s}_i\times 
( {\bf H}_{i,\rm tot} + \bbox{\zeta}_i )] 
- \gamma\lambda [{\bf s}_i\times [{\bf s}_i\times {\bf H}_{i,\rm tot} ] ]
\end{equation}
[cf. Eq. (\ref{langeq})], where $\bbox{\zeta}_i$ are postulated to be 
uncorrelated on different lattice sites, and
%
%
\begin{eqnarray}\label{htot}
&&
{\bf H}_{i,\rm tot} \equiv 
- \frac{1}{\mu_0}\frac{\partial {\cal H} }{ \partial {\bf s}_i }
\nonumber\\
&&\qquad
= {\bf H}_i + \frac{1}{\mu_0} \sum_j J_{ij}
( \eta_x {\bf s}_{xj} + \eta_y {\bf s}_{yj} + {\bf s}_{zj} ) 
\end{eqnarray}
is the total field acting on a given spin at the site $i$, 
which depends on 
the orientation of spins on the neighboring sites $j$.
In Eq. (\ref{htot}) 
${\bf s}_{\alpha j} \equiv s_{\alpha j}{\bf e}_\alpha$, 
$\alpha=x,y,z$, and ${\bf e}_\alpha$  
are the orts of the Descarte coordinate system.

The Fokker-Planck equation for the distribution function 
%
%
\begin{equation}\label{fsysdef}
f_{\rm tot}(\{{\bf N}_i\},t) = \bigg\langle \prod_{i=1}^{\cal N} 
\delta\bbox{(} {\bf N}_i - {\bf s}_i(t) \bbox{)} \bigg\rangle_\zeta
\end{equation}
of the whole system consisting of $\cal N$ spins can be derived in the same 
way as Eq. (\ref{fpe}) and has the form
%
%
\begin{eqnarray}\label{fpesys}
&&
\frac{\partial f_{\rm tot}}{\partial t} 
+ \sum_i \frac{\partial}{\partial {\bf N}_i}
\bigg\{
\gamma \left [{\bf N}_i\times {\bf H}_{i,\rm tot} \right] 
- \gamma\lambda [{\bf N}_i\times [{\bf N}_i\times {\bf H}_{i,\rm tot} ] ]
\nonumber \\
&&
\qquad
{} + \frac{\gamma\lambda T}{\mu_0} 
\left [{\bf N}_i\times 
\left [{\bf N}_i\times \frac{\partial}{\partial {\bf N}_i} 
\right ] 
\right ]
\bigg\} f_{\rm tot}= 0 .
\end{eqnarray}
One can check that the static solution of this equation is
%
%
\begin{equation}\label{ftotequi}
f_{{\rm tot},0}(\{{\bf N}_i\}) \propto 
\exp[-{\cal H}(\{{\bf N}_i\})/T]
\end{equation}
where $\cal H$ is given by Eq. (\ref{fham}).
Solving Eq. (\ref{fpesys}) is a formidable task that goes beyond the scope of 
this paper.
It is in any case not simpler than calculating averages with the 
distribution function (\ref{ftotequi}) at an equilibrium and requires 
application of some kind of many-body 
perturbation theory, as the diagram technique for classical spin systems
(see, e.g., Ref. \onlinecite{gar96prb}), 
which has proved to be rather efficient for description of their static
properties.
Here we resort to the mean field approximation with respect to spin-spin 
interactions, which means, however, dropping their contribution into the 
relaxation 
rates.
In MFA the distribution function of the system (\ref{fsysdef}) is 
multiplicative, and one can use the distribution functions $f_i$
for each spin on the site $i$,
which satisfy the Fokker-Planck equation (\ref{fpe}) with 
$H \Rightarrow H_i^{\rm MFA}$, where $H_i^{\rm MFA}$ is given by 
Eq. (\ref{htot}) with the replacement 
${\bf s}_i \Rightarrow {\bf m}_i \equiv \langle {\bf s}_i \rangle$.
Solution of such mean-field FPE's similar to that of Sec. \ref{secfpe} or
Sec. \ref{secsol} leads to the set of coupled LLB equations
for ${\bf m}_i$, $i=1,2,\ldots,{\cal N}$ of the type (\ref{llbm1}) in
a general nonlinear situation or Eq. (\ref{llbs}) for slow motions.
The static solution of these LLB equations satisfies the inhomogeneous 
Curie-Weiss equation,
%
%
\begin{equation}\label{cweiss}
{\bf m}_i = B(\xi_{0i}) \frac{ \bbox{\xi}_{0i} }{ \xi_{0i} },
\qquad \bbox{\xi}_{0i} \equiv \frac{ \mu_0 {\bf H}_i^{\rm MFA} }{ T },
\end{equation}
which describes within the MFA both the homogeneous state and such 
configurations as domain walls with account of thermal effects 
(see, e.g., Ref. \onlinecite{gar96jpa} and references therein).

For the most of ferromagnetic substances the small-anisotropy case,
i.e., $\eta'_{x,y} \equiv 1-\eta_{x,y} \ll 1$, is realized.
In this case the spatial inhomogeneity of magnetization at a distance of 
the lattice spacing is small, and one can use the continuous 
approximation.
For ${\bf H}_i^{\rm MFA}$ the latter means
%
%
\begin{eqnarray}\label{hmfa}
&&
{\bf H}^{\rm MFA}({\bf r}) \cong {\bf H}_E + {\bf H}'_{\rm eff},
\qquad {\bf H}_E = \frac{J_0}{\mu_0} {\bf m},
\nonumber\\
&&
{\bf H}'_{\rm eff} = {\bf H} + \frac{ J_0 }{ \mu_0 }
\big[ \alpha \Delta {\bf m} 
- \eta'_x {\bf m}_x - \eta'_y {\bf m}_y \big] ,
\end{eqnarray}
where $J_0$ is the zero Fourier component of the exchange interaction,
$\Delta$ is the Laplace operator, and $\alpha$ is a lattice-dependent 
constant 
(for the simple cubic lattice $\alpha=a_0^2/6$, 
where $a_0$ is the lattice spacing). 
The most important for ferromagnets is the case of the strong homogeneous 
exchange field,
$|{\bf H}_E| \gg |{\bf H}'_{\rm eff}|$, which is realized below 
$T_C=\frac{1}{3}J_0$, where there is a spontaneous magnetization, 
and also in the region just above $T_C$, where the longitudinal 
susceptibility is large.
As in this case the external field ${\bf H}(t)$ that can drive the system 
off the equilibrium is a relatively small quantity, one can use 
Eq. (\ref{llbm2}) [or, for slow motions, Eq. (\ref{llbs})] and expand 
${\bf m}_0 = B(\beta\mu_0 H^{\rm MFA}) {\bf H}^{\rm MFA}/H^{\rm MFA}$, where
$\beta\equiv 1/T$, up to the first order in ${\bf H}'_{\rm eff}$.
This leads to the equation 
%
%
\begin{eqnarray}\label{llbfm1}
&&
{\bf \dot m} = 
\gamma [{\bf m}\times {\bf H}'_{\rm eff}]
- \gamma \lambda_1
\left(
\frac{ 1 - B/m }{ \mu_0\beta B' }
- \frac{ {\bf mH}'_{\rm eff} }{ m^2 }
\right) {\bf m}
\nonumber\\
&&\qquad\qquad
- \gamma \lambda_2
\frac{ [{\bf m}\times [{\bf m}\times {\bf H}'_{\rm eff} ] ] }{m^2} , 
\end{eqnarray}
where $B=B(m\beta J_0)$,
%
%
\begin{equation}\label{lam1lam2}
\lambda_1 = 2\lambda \frac{T}{J_0}, \qquad 
\lambda_2 = \lambda \left( 1 - \frac{T}{J_0} \right),
\end{equation}
if Eq. (\ref{llbm2}) was used, and the same with 
$\lambda_1 \Rightarrow \lambda_1 \Gamma_{1,{\rm int}}/\Gamma_1$
for the ``slow'' LLB equation (\ref{llbs}).
The difference $1 - B/m$ in Eq. (\ref{llbfm1}) is a small quantity 
proportional to the deviation from the equilibrium.
It can be further simplified to
%
%
\begin{equation}\label{simpl}
\frac{ 1 - B/m }{ \mu_0\beta B' } \cong
\left\{
\begin{array}{ll}
\displaystyle
\frac{1}{2\tilde\chi_\|} \left( \frac{m^2}{m_e^2} - 1 \right),
& T<T_C \\
\displaystyle
\frac{J_0}{\mu_0} \left( \frac{3}{5} m^2 - \epsilon \right),
& |\epsilon|\ll 1 ,
\end{array}
\right.
\end{equation}
where $\epsilon \equiv 1 - T/T_C$, $m_e$ is the equilibrium spin 
polarization satisfying $m_e = B(m_e\beta J_0)$, and
%
%
\begin{equation}\label{tilsus}
\tilde\chi_\| = \frac{\partial m}{\partial H} = 
\frac{\mu_0}{J_0} \frac{B'\beta J_0 }{1-B'\beta J_0 }
\end{equation}
is the spin polarization susceptibility, calculated for $m=m_e$.
Using $B(\xi) \cong \frac{1}{3}\xi - \frac{1}{45}\xi^3 + \ldots$
and $m_e^2\cong \frac{5}{3} \epsilon$ near $T_C$,
one can check that the two expressions in Eq. (\ref{simpl}) overlap in this 
region.

The last step is to rewrite (\ref{llbfm1}) for the macroscopic 
magnetization, ${\bf M} = \mu_0 {\bf m}/v_0$, where $v_0$ is the unit-cell 
volume.
This leads to the final result
%
%
\begin{eqnarray}\label{llbfm2}
&&
{\bf \dot M} = 
\gamma [{\bf M}\times {\bf H}_{\rm eff}]
+ L_1 \frac{ ({\bf MH}_{\rm eff}){\bf M} }{ M^2 } 
\nonumber\\
&&\qquad
{} - L_2
\frac{ [{\bf M}\times [{\bf M}\times {\bf H}_{\rm eff} ] ] }{M^2} , 
\end{eqnarray}
where $L_1$ and $L_2$ 
are the longitudinal and transverse kinetic coefficients,
%
%
\begin{equation}\label{L1L2}
L_{1,2} = \gamma M_e \alpha_{1,2},
\qquad
\alpha_{1,2} = \lambda_{1,2}/m_e,
\end{equation}
$\alpha_1$ and $\alpha_1$ are the corresponding Gilbert damping parameters,
and the effective field ${\bf H}_{\rm eff}$ is given by 
%
%
\begin{eqnarray}\label{heff}
&&
{\bf H}_{\rm eff} =  
{\bf H} + \frac{1}{q_d^2}\Delta {\bf M} 
- \frac{1}{\chi_x}{\bf M}_x - \frac{1}{\chi_y}{\bf M}_y
\nonumber\\
&&\qquad
{} - \frac{1}{2\chi_\|} \left( \frac{M^2}{M_e^2} -1 \right) {\bf M} 
\end{eqnarray}
[cf. Eq. (\ref{hmfa})].
In Eq. (\ref{heff}) 
%
%
\begin{equation}\label{qdWd}
\frac{1}{q_d^2} \equiv \frac{\alpha J_0}{W_d},
\qquad W_d \equiv \frac{\mu_0^2}{v_0} ,
\end{equation}
$q_d$ and $W_d$ are the characteristic dipolar wave number and dipolar 
energy, $\alpha J_0$ is the second moment of the exchange interaction,
and the susceptibilities are given by
%
%
\begin{equation}\label{sus}
\chi_\| = \frac{W_d}{J_0}\frac{B'\beta J_0 }{1-B'\beta J_0 },
\qquad
\chi_{x,y} = \frac{W_d}{J_0} \frac{ 1 }{ 1-\eta_{x,y} }.
\end{equation}
The effective field ${\bf H}_{\rm eff}$ of Eq. (\ref{heff}) can be written as 
a variational derivative
%
%
\begin{equation}\label{heffvar}
{\bf H}_{\rm eff}({\bf r}) = \frac{ \delta F }{ \delta {\bf M}({\bf r}) },
\end{equation}
where $F$ is the MFA free energy of a ferromagnet,
%
%
\begin{eqnarray}\label{fenergy}
&&
F = F_0 + \int\!\! d{\bf r} 
\left\{
- {\bf HM}
+ \frac{1}{2q_d^2} (\nabla{\bf M})^2 
+ \frac{1}{2\chi_x} M_x^2
\right.
\nonumber\\
&&\qquad
\left.
{} + \frac{1}{2\chi_y} M_y^2
+ \frac{1}{8M_e^2\chi_\|} (M^2-M_e^2)^2
\right\} ,
\end{eqnarray}
$(\nabla{\bf M})^2\equiv (\nabla M_x)^2 + (\nabla M_y)^2 +
(\nabla M_z)^2$, and $F_0$ is the equilibrium
free energy in the absence of anisotropy and magnetic field.
The direct derivation of this free energy from the mean field theory 
is tricky and will be presented elsewhere.
Equation (\ref{fenergy}) provides a link between 
the ``micromagnetics'', \cite{lanlif35,bro63mm} which 
ignores changes of the magnetization 
magnitude $|{\bf M}|$, and the Landau theory of phase transitions, 
\cite{lan37,lanlif5}
which is a limiting form of the MFA 
pretending to be valid only in the vicinity of $T_C$ 
where the order parameter ${\bf M}({\bf r})$ is small.
In fact, for weakly anisotropic systems in a magnetic field smaller than 
the homogeneous exchange field $H_E$, the actual small quantity,
which remains small in the whole temperature range,
is not $M^2({\bf r})$, but rather the difference 
$M^2({\bf r}) - M_e^2$, 
where $M_e$ is the equilibrium 
magnetization in the absence of anisotropy and magnetic field.
Since in the MFA near $T_C$ one has 
$M_s^2 \propto \chi_\|^{-1} \propto \epsilon \equiv 1-T/T_C$,
the last term of Eq. (\ref{fenergy})
takes on the Landau form $AM^2 + BM^4$ with $A=-\epsilon A_0$, and 
$A_0, B = {\rm const}$.
This shows, further, that Eq. (\ref{fenergy}) can be continued into the region
$T>T_C$ as the usual Landau theory.
The free energy Eq. (\ref{fenergy}) can be brought into the ``micromagnetic'' 
form by introducing the magnetization direction vector
$\mbox{\boldmath$\nu$} \equiv {\bf M}/M$.
One can then identify
%
%
\begin{equation}\label{identmicromag}
\frac{1}{2\chi_{x,y}} M_{x,y}^2 = K_{x,y} \nu_{x,y}^2,
\qquad
K_{x,y} = \frac{M^2}{2\chi_{x,y}},
\end{equation}
where $K_{x,y}$ are the anisotropy constants.

\section{Discussion}
\label{secdisc}

In this paper several forms of the Landau-Lifshitz-Bloch (LLB) equation of 
motion for a single classical spin interacting with the bath as well as 
for classical ferromagnets within the MFA have been obtained.
These LLB equations are applicable for all temperatures and 
contain both transverse and longitudinal relaxation terms.
The {\em nonlinear} response of a single spin to the arbitrary changing 
magnetic field ${\bf H}(t)$ is the most accurately described by the 
nonlinear LLB equation (\ref{llbm1}).
For slowly varying ${\bf H}(t)$ the exact ``slow'' LLB equation 
(\ref{llbs}) containing the integral longitudinal relaxation time 
Eq. (\ref{tauint}) can be used.
This case is the most important one for the domain-wall dynamics.
For ferromagnets within the MFA the magnetic field ${\bf H}$ 
in the LLB equation should be replaced by ${\bf H}^{\rm MFA}$, 
which is given by Eq. (\ref{htot}) with 
${\bf s}_i \Rightarrow {\bf m}_i \equiv \langle {\bf s}_i \rangle$ 
in a general case or by Eq. (\ref{hmfa}), if the continuous approximation is 
applicable.
If, additionally, in Eq. (\ref{hmfa}) the homogeneous exchange field 
${\bf H}_E$ is dominant, which is typical for ferromagnets below 
and near above $T_C$, the LLB equation reduces to the form 
(\ref{llbfm2}) with Eqs. (\ref{heffvar}) and (\ref{fenergy}).
Equation (\ref{llbfm2}) could be also written, without specifying the 
form of kinetic coefficients $L_{1,2}$ and that of the free energy 
Eq. (\ref{fenergy}), from general arguments.
It very close to the phenomenological Bar'yakhtar
equation, \cite{bar84,bar87} which contains an additional relaxation term
proportional to $\Delta H_{\rm eff}$.
This term, whose microscopic origin is the 
spin-spin interaction or the correlation of the Langevin fields 
$\bbox{\zeta}_i$ in Eq. (\ref{fmlleq}) on different lattice sites $i\ne j$,
was shown, \cite{barivasaf89} however, to yield
a contribution into the domain-wall dynamics, which is negligible in 
comparison to that of the longitudinal relaxation term in Eq. (\ref{llbfm2}).

The quantum generalization of the nonlinear LLB equation (\ref{llbm1}),
which contains additional relaxation terms of a different symmetry,
was derived in Ref. \onlinecite{gar91llb} by the approximate solution 
of the density matrix equation for a single spin interacting with an 
idealized phonon bath, which is based on choosing the 
distribution function of the type similar to Eq. (\ref{fmod}).
In the classical limit the density-matrix equation goes over to the FPE 
and, accordingly, the quantum LLB simplifies to Eq. (\ref{llbm1}) with 
the microscopically determined bath-coupling parameter $\lambda$.
For ferromagnets with the dominant homogeneous exchange interaction the 
quantum LLB equation simplifies to the same macroscopic form 
(\ref{llbfm2}).
The main result of the present paper --- the ``slow'' LLB equation 
(\ref{llbs}) --- can be obtained in the quantum case, too, by a 
perturbative solution of the density-matrix equation for a slowly changing 
magnetic field, which is similar to the derivation in Sec. \ref{secsol}.
The final result can be, however, obtained by replacing 
$\Gamma_1 \Rightarrow \Gamma_{1,{\rm int}}\equiv \tau_{\rm int}^{-1}$ 
in the longitudinal relaxation term.
The analytical expression for $\Gamma_{1,{\rm int}}$ in the quantum case 
without 
single-site anisotropy was already given in Ref. \onlinecite{gar91llb}.
Very recently it was generalized for the anisotropic case to describe the 
thermoactivation escape rate of quantum spin systems. \cite{gar97pre}

The most serious problem by the derivation of the LLB equation for 
ferromagnets, that has not been solved yet, is taking into account the 
spin-spin interactions.
This is a rather difficult task, since one should consider the FPE 
(\ref{fpesys}) for the whole system, which describes all possible static 
and dynamic spin correlations.
Even at an equilibrium, where the solution of the FPE (\ref{fpesys}) 
is known and given by Eq. (\ref{ftotequi}), 
one faces the problem of a phase transition in a many-body system.
Calculation of spin-spin contributions into the longitudinal and transverse 
kinetic coefficients $L_1$ and $L_2$ in the LLB equation for 
ferromagnets, Eq. (\ref{llbfm2}), goes beyond the scope of this paper and is 
planned for the future.

\section*{Acknowledgments}

The author thanks Hartwig Schmidt for valuable discussions.
The financial support of Deutsche Forschungsgemeinschaft 
under Contract No. Schm 398/5-1 is greatfully acknowledged.

\appendix

\section*{Derivation of the Fokker-Planck equation}

Here the derivation of the Fokker-Planck equation (\ref{fpe}) is 
presented, which is more direct and simple than the original one by 
Brown \cite{bro63} and which uses more advanced stochastic methods 
applied, in particular, in the dynamical renormalization-group (RG) theory. 
\cite{mamaz75,hohhal77,zin89}
The RG considerations start, however, with ``soft-spin'' models with  
the formal Langevin {\em sources} 
(i.e., the inhomogeneous terms in the stochastic differential equations 
for the spin density), which cannot be interpreted as  
random {\em magnetic fields} acting on spins.
For our purposes, we will derive the FPE 
for magnetic systems with the methods of Refs. \onlinecite{mamaz75,zin89}
but starting from the more 
realistic stochastic Landau-Lifshitz equation (\ref{langeq}).
At first we introduce the probability distribution of the random 
Gaussian noise $\bbox{\zeta}$,
%
%
\begin{equation}\label{zdistr}
{\cal F}[\bbox{\zeta}(\tau)] = \frac{ 1 }{ {\cal Z}_\zeta }
\exp \left[
- \frac{1}{2a}\int_{-\infty}^\infty\!\! d\tau 
\bbox{\zeta}^2(\tau) 
\right] ,
\end{equation}
where ${\cal Z}_\zeta = \int\!\! D\bbox{\zeta}\,{\cal F}$ 
is the noise partition function, $\int\!\! D\bbox{\zeta}$ denotes 
functional integration over
realizations of $\bbox{\zeta}(\tau)$ and 
$a \equiv 2\lambda T/(\gamma\mu_0)$.
With the help of Eq. (\ref{zdistr}) the average of any noise functional 
${\cal A}[\bbox{\zeta}]$ can be written as
%
%
\begin{equation}\label{aavr}
\langle {\cal A}[\bbox{\zeta}] \rangle_\zeta =
\int\!\!\! D\bbox{\zeta}\, {\cal A}[\bbox{\zeta}] {\cal F}[\bbox{\zeta}] .
\end{equation}
With the use of the obvious identity
%
%
\begin{equation}\label{zvar}
\frac{ \delta \zeta_\alpha(\tau) }{ \delta \zeta_\beta(t) } =
\delta_{\alpha\beta} \delta(\tau-t)
\end{equation}
one can calculate variations of ${\cal F}[\bbox{\zeta}]$ of Eq. (\ref{zdistr}):
%
%
\begin{eqnarray}\label{fvar}
&&
\frac{ \delta {\cal F}[\bbox{\zeta}] }{ \delta \zeta_\alpha(t) } =
- \frac{1}{a} \zeta_\alpha(t) {\cal F}[\bbox{\zeta}] ,
\\
&&
\frac{ \delta^2 {\cal F}[\bbox{\zeta}] }
{ \delta \zeta_\alpha(t) \delta \zeta_\beta(t') } =
\left[
\frac{1}{a^2} \zeta_\alpha(t) \zeta_\beta(t')
%
- \frac{1}{a} \delta_{\alpha\beta} \delta(t-t')
\right]
{\cal F}[\bbox{\zeta}] ,
\nonumber
\end{eqnarray}
etc.
Since for all $n$ one has
%
%
\begin{equation}\label{varint}
\int\!\!\! D\bbox{\zeta}\,
\frac{ \delta^n {\cal F}[\bbox{\zeta}] }
{ \delta \zeta_{\alpha_1}(t_1) \delta \zeta_{\alpha_2}(t_2) \ldots
\delta \zeta_{\alpha_n}(t_n)} = 0 ,
\end{equation}
the functional integration of Eq. (\ref{fvar}) leads to 
$\langle \zeta_\alpha(t)\rangle=0$ and Eq. (\ref{zetacorr}).
Further, one can show that all averages of an odd number of $\bbox{\zeta}$ 
components are zero and those of an even number $n>2$ of $\zeta$'s decay 
pairwise and can be expressed through the pair average Eq. (\ref{zetacorr}), 
i.e., the statistics of the random field $\bbox{\zeta}(t)$ is Gaussian.

The distribution function of spins $f$ is determined as
%
%
\begin{equation}\label{fdef}
f({\bf N},t) \equiv \langle \pi(t,[\bbox{\zeta}]) \rangle_\zeta, 
\qquad
\pi(t,[\bbox{\zeta}]) \equiv \delta \bbox{(}{\bf N }-{\bf s}(t) \bbox{)} .
\end{equation}
The time derivative of $f$ can be calculated using 
%
%
\begin{equation}\label{pider}
{\bf \dot \pi} = - \frac{\partial \pi }{ \partial {\bf N} } {\bf \dot s}
\end{equation}
and the equation of motion (\ref{langeq}), which yields
%
%
\begin{eqnarray}\label{fder}
&&
\frac{\partial f}{\partial t} =
- \frac{\partial}{\partial {\bf N}}
\bigg\{
\gamma [ {\bf N} \times {\bf H} ] f 
- \gamma\lambda [ {\bf N} \times [ {\bf N} \times {\bf H} ] ] f
\nonumber \\
&&
\qquad
{} + \gamma [ {\bf N} \times \langle \bbox{\zeta}(t)\pi(t,[\bbox{\zeta}]) 
\rangle_\zeta ] \bigg\} .
\end{eqnarray}
Then the average 
$\langle \bbox{\zeta}(t)\pi(t,[\bbox{\zeta}]) \rangle_\zeta$
can be transformed with the use of the first of Eqs. (\ref{fvar}) 
and integration by parts,
%
%
\begin{eqnarray}\label{trans}
&&
\langle \bbox{\zeta}(t) \pi(t,[\bbox{\zeta}]) \rangle_\zeta =
- a\!\! \int\!\!\! D\bbox{\zeta}\, \pi(t,[\bbox{\zeta}]) 
\frac{ \delta {\cal F}[\bbox{\zeta}] }{ \delta \bbox{\zeta}(t) }
\nonumber\\
&&\qquad
= a \bigg\langle \frac{ \delta \pi(t,[\bbox{\zeta}]) }
{ \delta \bbox{\zeta}(t) } \bigg\rangle
= - a \bigg\langle \frac{ \partial \pi }{ \partial N_\beta }
\frac{ \delta s_\beta(t,[\bbox{\zeta}]) }
{ \delta \zeta_\alpha(t) } \bigg\rangle {\bf e}_\alpha ,
\end{eqnarray}
where ${\bf e}_\alpha$ with $\alpha=x,y,z$ 
are the orts of the Descarte coordinate system 
and summation over components $\alpha,\beta$ is implied.
The variational derivative $\delta s_\beta/\delta \zeta_\alpha$ can be 
calculated, if one writes down the formal solution of the 
stochastic Landau-Lifshitz equation (\ref{langeq}),
%
%
\begin{equation}\label{formsol}
s_\beta(t) = \gamma \int_{t_0}^t\!\! dt' e_{\beta\gamma\alpha}
s_\gamma(t') [H_\alpha(t')+\zeta_\alpha(t')] + \ldots ,
\end{equation}
where $e_{\beta\gamma\alpha}$ is the antisymmetric unit tensor.
One can see that
%
%
\begin{equation}\label{svar}
\frac{ \delta s_\beta(t,[\bbox{\zeta}]) }{ \delta \zeta_\alpha(t') } =
\left\{
\begin{array}{ll}
\gamma e_{\beta\gamma\alpha} s_\gamma(t'), & t'<t \\
0                                          & t'>t .
\end{array}
\right.
\end{equation}
For $t=t'$ the above calculation does not yield a definite value of 
$\delta s_\beta/\delta \zeta_\alpha$, but with the help of the usual 
arguments based on the regularization of $\delta$ functions \cite{zin89}
the latter can be found to be 
$\frac{1}{2}\gamma e_{\alpha\beta\gamma} s_\gamma(t)$.
Now Eq. (\ref{trans}) can be finally written in the form
%
%
\begin{equation}\label{diffus}
\langle \bbox{\zeta}(t) \pi(t,[\bbox{\zeta}]) \rangle_\zeta =
\frac{\gamma a}{2}
\left[ {\bf N} \times \frac{\partial f}{\partial {\bf N}} \right] .
\end{equation}
Adopting it in Eq. (\ref{fder}), one comes to the Fokker-Planck equation 
(\ref{fpe}).

\end{document}